\documentclass{PoS}

\newcommand{\met}       {\mbox{$\not\!\!E_T$}}

\title{Top Quark Mass Measurements at the Tevatron}

\ShortTitle{Top Quark Mass Measurements at the Tevatron}

\author{\speaker{Reinhild Yvonne Peters}\thanks{on behalf of the CDF and D0 Collaborations.}\\
        Georg-August Universit\"at G\"ottingen, also at DESY\\
        Friedrich-Hund-Platz 1 \\
        37077 G\"ottingen\\ 
        Germany \\
        E-mail: \email{repeters@cern.ch}}

\abstract{Since the discovery of the top quark in 1995 by the CDF and
  D0 collaborations at the Fermilab Tevatron proton antiproton collider, 
precise measurements of its mass are ongoing. Using data recorded by
the D0 and CDF experiment, corresponding to up to the full Tevatron data sample, top quark mass measurements performed in different final states using various extraction techniques are presented in this article. 
The recent Tevatron top quark mass combination yields $m_t=173.20\pm 0.87$~GeV.
Furthermore, measurements of the  top antitop quark mass difference from the Tevatron are discussed. }

\FullConference{The European Physical Society Conference on High Energy Physics -EPS-HEP2013\\
		18-24 July 2013\\
		Stockholm, Sweden}

\begin{document}

\section{Introduction}
In 1995, top quark  was  independently   discovered by the CDF and D0 collaborations at the Tevatron 
proton antiproton ($p\bar{p}$)
collider at Fermilab~\cite{cdfobs,d0obs}. It is the
heaviest known elementary particle. The top quark mass, $m_t$, is a free
parameter in the standard model of particle physics (SM). Especially
in the light of discovery of a Higgs boson by the ATLAS and CMS
experiments at the LHC~\cite{HiggsAtlas,HiggsCMS}, it is
important to  measure the top quark mass as precisely as possible and
test its role in the mechanism of electroweak symmetry breaking. 

Since the discovery of the top quark, several methods to  measure the top
quark mass have been developed and refined. Furthermore, various $t\bar{t}$ final
states have been explored for these measurements. Thanks
to dedicated methods and the large data samples of CDF and D0,
precision measurements of the top quark mass in several
channels exist by now. 
In order to provide the most precise number for the top quark mass, the individual measurements
from both CDF and D0 were combined into one value.

In the following, the recent Tevatron combination of direct top quark
mass measurements is presented, as well as recent results from the CDF and
D0 collaborations that serve as inputs to the combination. In addition, an
alternative approach to extract the top quark mass from the measurement
of the $t\bar{t}$  cross section and  measurements
of the top antitop quark mass difference are discussed.

\section{Direct Top Quark Mass Measurements and Tevatron Combination}
The main production mechanism of top quarks at the Tevatron is via top
quark pair production ($t\bar{t}$). In the SM, top quarks decay with
almost 100\% probability into a $b$~quark and a $W$~boson. The final
states are classified according to the decays of the two $W$~bosons
from $t\bar{t}$. The main channels are the dileptonic final state with
both $W$~bosons decaying into leptons, the lepton+jets channel with
one $W$~boson decaying into leptons and the other into a pair of
quarks, and the all hadronic final state where both $W$~boson decays
hadronically.  Channels where at least one $W$~boson decays
into a hadronically decaying tau-lepton are not considered in the 
analyses discussed here.

The recent combination of top quark mass measurements from the
Tevatron uses results from Run~I (1.8~TeV collision energy) and Run~II
(1.96~TeV collision energy) of the Tevatron in
different final states and using different extraction methods, where
data samples corresponding to an integrated luminosity of up to 8.7~fb$^{-1}$ were analyzed. The combination is
performed using the Best Linear Unbiased Estimator (BLUE), taking into
account correlations between systematic uncertainties. In the BLUE
method, each input measurement receives a weight with which it
contributes to the combination. The combined value of 
the top quark mass is  $m_t=173.20\pm0.51{\rm (stat)}
\pm0.71{\rm(syst)}$~GeV~\cite{tevatroncombi}. This result is limited
by systematic uncertainties, where the dominant ones arise from signal
modeling and the calibration of the light jet energy scales. The
$\chi^2$ of the combination is 8.5 for 11 degrees of freedom, which
corresponds to a probability of 76\% that the results are consistent
between each other. 


The result with the  highest weight in the combination is  the  lepton+jets analysis by CDF, which uses the full Run~II data sample of
8.7~fb$^{-1}$~\cite{cdftopmassljets}. In the lepton+jets final state
one $W$~boson from top quark decays into a charged lepton and a
neutrino, which leaves the detector without interacting. The other $W$
boson decays into a pair of quarks. The $t\bar{t}$ kinematics are
reconstructed using a kinematic fit, where constraints from the
known $W$~boson mass as well as the requirement that the reconstructed
top and antitop masses are the same are included. An unbinned maximum
likelihood fit is
then performed using the reconstructed top quark mass $m_t^{reco}$
with lowest and second lowest $\chi^2$ value. To enhance the
sensitivity, events with no, one or two identified $b$-jets are
treated separately. The influence of the jet energy scale (JES) uncertainty
on the result of the top quark mass measurement is reduced by fitting
the JES in-situ using the two quarks from the $W$~boson decay and
constraining their invariant mass to the knwon $W$~boson mass. The
measured top quark mass, $m_t= 172.85\pm0.71{\rm (stat)}\pm0.85{\rm
  (syst)}$~GeV representes the best single measurement to date. The
systematic uncertainties are dominated by effects from the choice of
Monte Carlo (MC) generator, color reconnection, the residual JES
uncertainty and the $b$-jet energy scale. 

The result with the second highest weight in the Tevatron combination
is coming from  an analysis in the lepton+jets final state by D0, using the
matrix element (ME) method on 3.6~fb$^{-1}$ of Run~II
data~\cite{d0meljets}. The ME method explores the full
kinematic information of each event by calculating per-event
signal probabilities $P_{sig}(x;m_t)$ and background probablities $P_{bkg}(x)$, where $x$ denotes the momenta
of the final state partons. The probabilities are calculated by
integrating over the leading
order (LO) matrix element for the $t\bar{t}$ production, folded with
the parton distribution functions and transfer functions, which 
describe  the transition of
the parton momenta into the measured
momenta of the final state particles from the top quark decays.
The measured top
quark mass is then obtained by maximizing the likelihood of the
product of these per-event probabilities.
Constraining the JES via the hadronically decaying $W$~boson, the
measured value is $m_t= 174.94\pm0.83{\rm (stat)}\pm0.78{\rm (JES)}
\pm 0.96{\rm
  (syst)}$~GeV. The dominant systematic uncertainties arise from
the underlaying event and hadronization and color
reconnection affecting the signal modeling, and from jet energy resolution and jet response
uncertainties.  

A new analysis by CDF using events with missing transverse energy
(\met) and jets on 8.7~fb$^{-1}$ of data represents the analysis
having the third highest weight in the Tevatron combination. 
 In this analysis, mainly lepton+jets events, where the electron or
 muon is not reconstructed, are recovered. The mass is extracted using
 templates of $m_t^{reco}$ with highest and second highest $\chi^2$,
 where the $\chi^2$ values are calculated using a modified kinematic fitter
 which only uses jets and the \met. The additional assumption in this
 kinematic fitter  with respect to the kinematic
 fitter in lepton+jets events as used in~\cite{cdftopmassljets} is that now both decay particles of the leptonically
 decaying $W$~boson are missing. Events with four, five or six jets
 are treated seprately, as are events with one or more than one
 identified $b$-jet. In events with five jets, $\tau$+jets events are
 assumed, where the $\tau$ is misidentified as jet and the kinematic
 fitter is adjusted accordingly. 
The top quark mass yields $m_t=173.93\pm 1.64{\rm (stat)}\pm0.87{\rm
  (syst)}$~GeV~\cite{cdfmetjets}, where the main contribution to the
systematic uncertainties comes from effects on the residual JES and
the signal modeling due to the choice of MC generator.

The next highest weight is provided by a measurement in the full
hadronic final state by CDF, using 5.8~fb$^{-1}$ of data. For this
analysis events with at least six jets, no leptons and no significant
\met are
used. Among the six leading jets, one or at least one have to be
identified as $b$-jets. 
Templates of $m_t^{reco}$ are constructed via a $\chi^2$ like quantity. The two
hadronically decaying $W$~bosons are used to constrain the JES. The
main challenge in this analysis is the large background from QCD
(quantum chromo dynamics)
multijet events. The background consists of the QCD
production of light and heavyflavor quarks, which is hard to
simulate. Therefore this background is constructed using a data-driven
method, which is based on the
parametrization of the rate with which to identify a $b$-jet, derived
in a sample of events with five jets that is dominated by
background. Before extracting the mass, the sample is enriched in $t\bar{t}$ events, a neural
network is constructed, that is based on variables that depend on the
energy, direction and shape of the jets. The fitted top quark mass is
$m_t=172.5\pm 1.4{\rm (stat)}\pm 1.4{\rm
  (syst)}$~GeV~\cite{cdfallhadronic}. The main contribution to the
systematic uncertainties arises from the choice of MC generator for
the modeling of the signal, the construction of the QCD multijet
background and uncertainties on the residual JES.

\begin{figure*}[t]
\centering
\includegraphics[width=60mm]{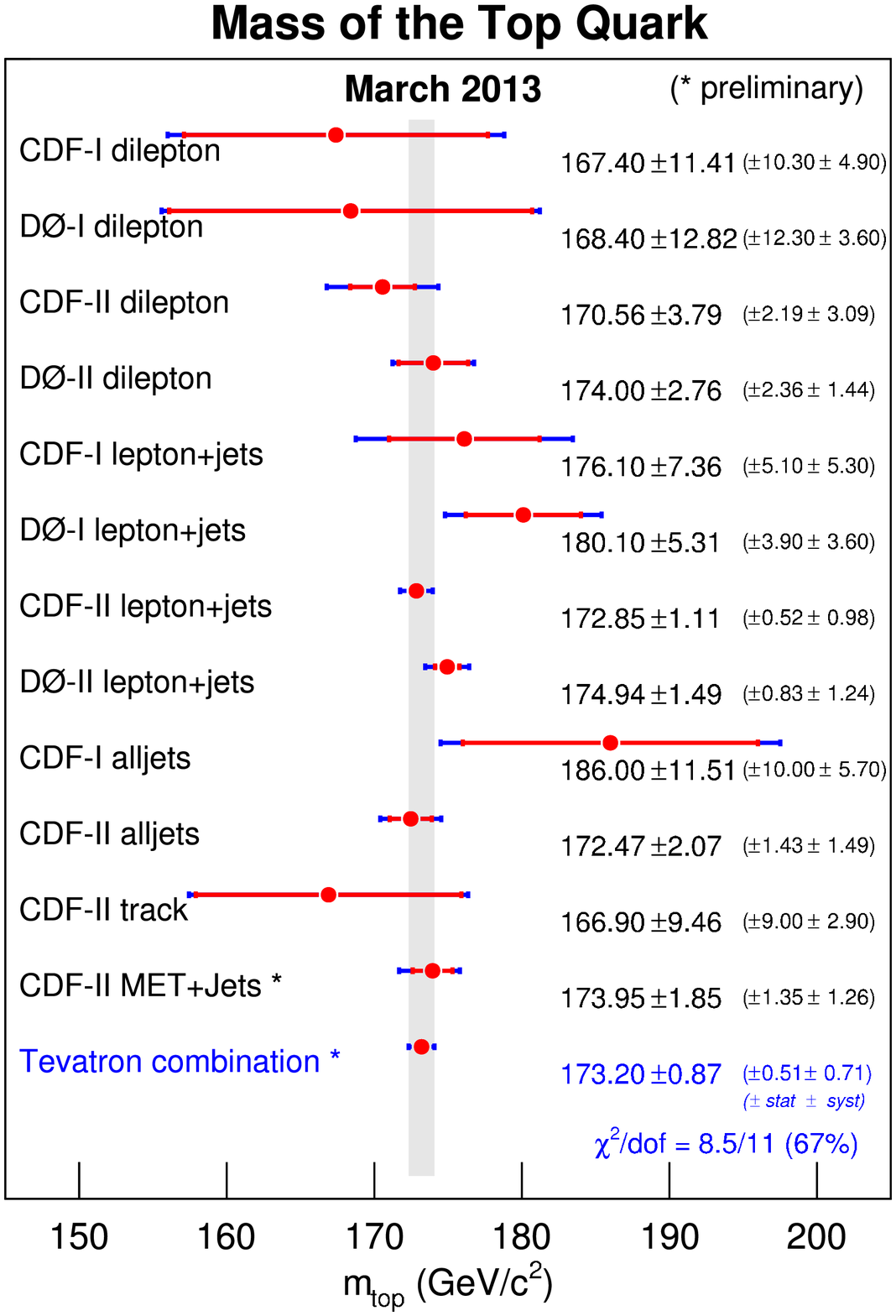}
\includegraphics[width=88mm]{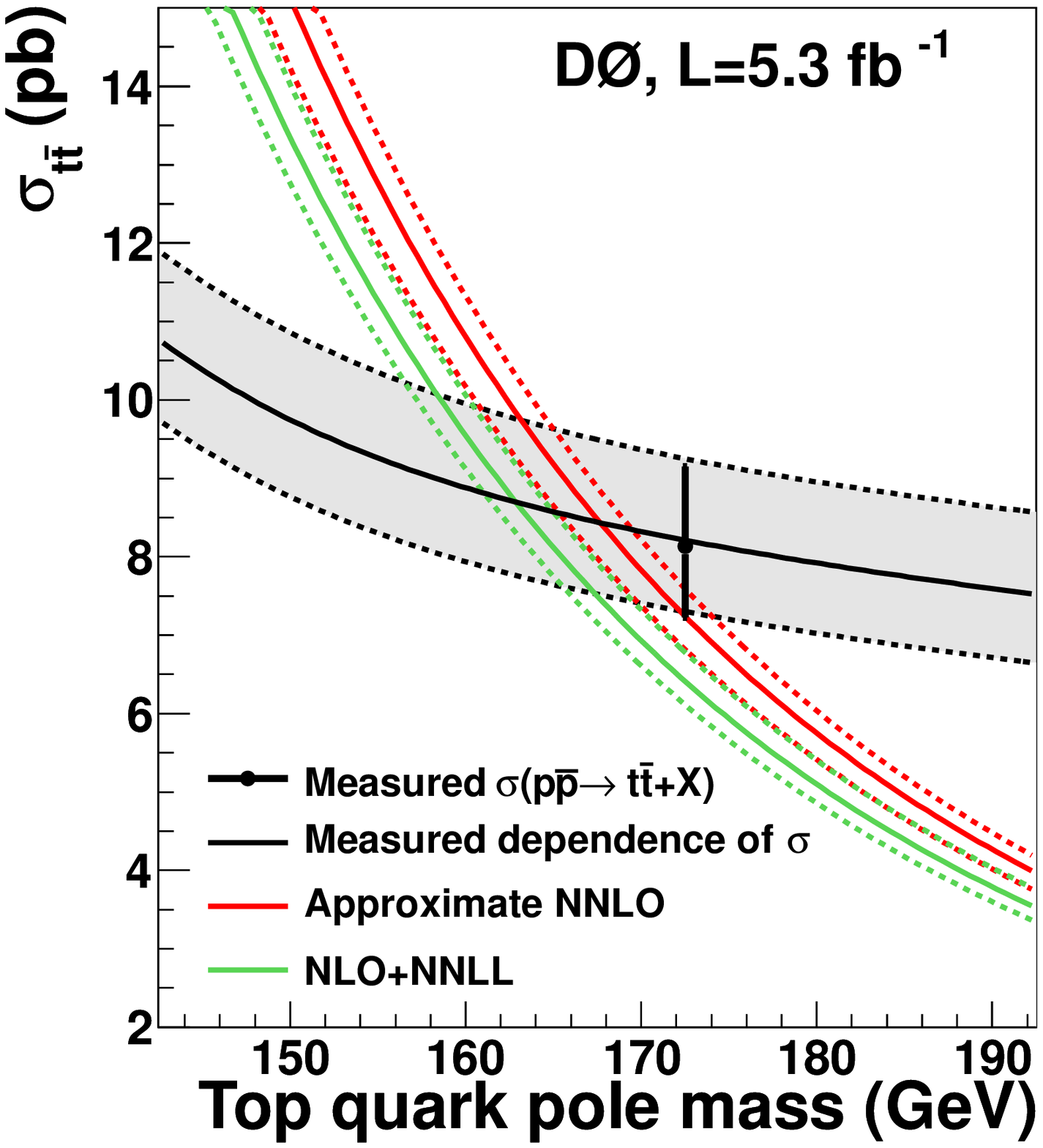}
\caption{ Recent Tevatron combination of top quark mass measurements
  from CDF and D0 (left)~\cite{tevatroncombi} and top quark mass extraction from the
  $t\bar{t}$ cross section (right)~\cite{d0massfromxsec}.} \label{fig:combi}
\end{figure*}

Another important contribution to the Tevatron combination is provided
by the measurement of the top quark mass in the dileptonic final
state, performed by D0 on 5.3~fb$^{-1}$ of data. The dilepton final
state, where two isolated charged leptons (electron or muon) with
large transverse momentum are required, has the advantage of being a
very clean signature with small contribution from backgrounds. The
main  challenge is the more difficult reconstruction of the $t\bar{t}$
event from the final state objects, due to the presence of the two
neutrinos. The latter results in an under-constrained system. At D0,
two methods were explored in the recent measurement of the top quark
mass, where once the ME method was explored~\cite{d0dilepme}, and in the other analysis
the neutrino weighting method. The neutrino weighting technique uses
assumptions about the $\eta$~\footnote{The rapidity $y$ and pseudorapidity $\eta$ are defined as functions
    of the polar angle $\theta$ and parameter $\beta$ as
    $y(\theta,\beta) \equiv                                                                                                                                                                                                                                                            
    {\frac{1}{2}} \ln{[(1+\beta\cos{\theta})/(1-\beta\cos{\theta})]}$ and
    $\eta(\theta) \equiv y(\theta,1)$, where $\beta$ is the ratio of a particle's
    momentum to its energy. } of both neutrinos. For each neutrino
  $\eta$ sampling, solutions for the kinematics of the event are
  calculated, and a weight for each solution is assigned which is based on the agreement between the calculated neutrino transverse
momenta and the  measured value of the missing
transverse energy. For the measurement of the top quark mass, the
first two moments of the weight distribution are used to extract  $m_t$. In addition, this analysis uses the measurement of the
in-situ JES from the top quark mass measurement in the  lepton+jets
final state~\cite{d0meljets} to constrain the JES in the dilepton
channel. This provides an improved JES calibration in the
analysis. The $m_t$ measurements with both methods have been combined
using the BLUE method, resulting in $m_t=173.9\pm 1.9{\rm (stat)}\pm 1.6{\rm
  (syst)}$~GeV~\cite{d0dilep}. The main systematic uncertainties arise
from uncertainties on the JES and the  choice of MC generator for the
modeling of the signal. Figure~\ref{fig:combi} (left) shows the different channels contributing to the combination as well as the combined value itself.

\section{Other Top Quark Mass related Measurements}
The methods discussed in the previous section of direct top quark mass measurements rely on MC
simulation. Until today, it is under theoretical investigation how the measured top
quark mass from MC is related to  top quark pole or ${\overline {MS}}$
mass. The D0 collaboration has performed a determination of the top
quark mass from the measurement of the $t\bar{t}$ cross section using
5.3~fb$^{-1}$~\cite{d0massfromxsec}, which allows an unambiguous
interpretation of the top quark mass. In this analysis the measured
$t\bar{t}$ cross section is compared to inclusive calculations as
function of the top quark mass. Using the pole mass for the inclusive cross section calculations,
D0 extracted a pole mass of $m_t=167.5^{+5.2}_{-4.7}$~GeV for the
cross section calculation from Ref.~\cite{mochuwer}, compatible with
the top quark mass value from the Tevatron combination. Doing the same
extraction again but with a calculation in the ${\overline {MS}}$ mass
scheme yields about 7~GeV smaller values for
$m_t$. Figure~\ref{fig:combi} (right) shows the measured versus
calculated $t\bar{t}$ cross section as function of the top quark mass.

The direct top quark mass measurements assume the top and the antitop
to be equally heavy. A mass difference between particle and its
antiparticle would indicate CPT violation. 
The D0 and CDF collaborations have performed measurements of the top
antitop quark mass difference. 
The first measurement of  the mass difference between a bare quark
and its antiquark was performed by
the D0 collaboration using the ME method using 1~fb$^{-1}$ of
data in the lepton+jets final state~\cite{firstmassdiffd0}. By
repeating the measurement on 3.6~fb$^{-1}$ of data, D0 has extracted 
$m_t-m_{\bar{t}}=0.8 \pm 1.8 {\rm (stat)} \pm 0.5 {\rm                                                                                                                                                                                                                                 
  (syst)}$~GeV~\cite{massdiffd0}.
The
CDF collaboration performed the mass difference measurement using a
template technique in the lepton+jets channel, first using
5.6~fb$^{-1}$ of data~\cite{massdiffcdf}, and then repeating the same measurement on the
full Tevatron data sample of 8.7~fb$^{-1}$, resulting in
$m_t-m_{\bar{t}}=-1.95\pm1.26{\rm
  (stat+syst)}$~GeV~\cite{cdffinalmassdifference}. All results are
compatible with the SM prediction and are limited by the statistical
uncertainty.

\section{Conclusion and Outlook}
The measurement of the top quark mass is one of the legacy
measurements at the Tevatron. 
The most recent top quark mass combination from the Tevatron has been
presented in this article, as well as recent analyses in various channels that serve
as input to the combination. Furthermore, the determination of the top
quark mass from the $t\bar{t}$ cross section and measurements of the 
top antitop quark mass difference have been discussed. Some of the presented
measurements already use the full Tevatron data sample. Currently,
many different analyses in various channels are still ongoing to
perform the final top quark mass  measurements at D0 and CDF. Besides
analysing the full data sample, the main challenge for top quark mass
measurements at the Tevatron and the LHC is the understanding of the
dominant sources of systematic uncertainty, mainly involving studies
on the modeling of the $t\bar{t}$  signal. Given the importance of the
top quark mass for the understanding of the SM, in
particular in the context of electroweak symmetry breaking, the
precise measurement of the top quark mass using various methods will
stay an interesting field of study at all collider experiments.

\end{document}